\documentstyle[11pt,paspconf,twoside,epsf]{article}

\begin{document}

\title{Evidence for dark matter in different scales from the 
	KLUN galaxy sample} 

\author{M. O. Hanski, P. Teerikorpi and T. Ekholm}
\affil{Tuorla Observatory, Piikki\"{o}, Finland}

\author{G. Theureau}
\affil{Osservatorio Astronomico di Capodimonte, Naples, Italy}

\author{Yu. Baryshev}
\affil{Astronomical Institute of St.Petersburg University, St.Petersburg, 
	Russia}

\author{G. Paturel and P. Lanoix}
\affil{Observatoire de Lyon, Lyon, France}

\section{Introduction}
The KLUN (Kinematics of the Local UNiverse)
sample of 6600 spiral galaxies is used in studying 
dark matter in different scales:
\begin{itemize}
\item Type dependence of the zero-point of Tully-Fisher 
relation indicates $M/L \approx$ 9 -- 16 in galactic scale
\item Preliminary results from a study of selection effect influencing
double galaxies
give a larger value $M/L \approx$ 30 -- 50
\item Study of the Perseus-Pisces supercluster, using Malmquist
bias corrected TF distances and Tolman-Bondi solutions
indicates $M/L \approx$ 200 -- 600 for the large clusters. Similar
results were obtained in our previous work on Virgo galaxies
\item Application of developed version of Sandage-Tammann-Hardy
test of the linearity of Hubble law inside the observed
hierarchical (fractal) galaxy distribution up to 200 Mpc
suggests that either $\Omega_0$ is very small (0.01)
or major part of the matter is uniformly distributed dark matter.
\end{itemize}

\section{KLUN}

The KLUN sample currently contains 6600 spiral galaxies having measurements
of isophotal diameter ($D_{25}$), HI line width, radial velocity, 
and B-magnitude.
The sample is selected according to apparent diameter and is complete
downt to $D_{25}=1.6$ arcmin. Morphological type range from Sa to Sdm
is covered ($T=$ 1--8). The data are extracted from the homogenized
compilation catalogue LEDA (Lyon-Meudon Extragalactic database, 
http://www-obs.univ-lyon1.fr/leda) and complemented with redshift and
HI spectrum measurements. Previously the KLUN sample has been used for
the determination of the Hubble parameter with the Tully-Fisher (TF) relation.
After taking properly into account the Malmquist selection and calibrator
sample biases, and using the type dependent TF relation we obtained 
$H_0 \approx 55$ km~s$^{-1}$~Mpc$^{-1}$ with both the direct (Theureau et al.\
1997b) and the inverse (Ekholm et al.\ 1999) TF relation.

\section{Type dependence in the TF relation and disk + bulge + dark halo model}

The revealed Hubble type dependence in the zero-point of the TF relation
has been interpreted with a simple mass model (exponential disk + 
bulge + dark halo $\propto$ luminous mass, ref.\ Theureau et al.\ 1997a).
The fraction of dark mass was found to be in the range 50 -- 80 \% within
the radius that the TF measurements refer to. This corresponds to $M/L
\approx$ 9 -- 16, depending on the type. A better evaluation is expected when
one takes into account a luminosity dependent $M/L$ ratio (Hanski \&
Teerikorpi, in prep.).
\begin{table}
\caption[]{The simple model described in Theureau et al.\ (1997a)
uses a common $M/L = 3.72$ for the bulge, and type dependent $M/L$ value for
the disk component of spirals. Adding the gas masses and assuming 
constant dark mass fraction the
total $M/L$ of spirals can be estimated by fitting the observed TF zero points
to the ones predicted by this model. The values below are from Table 3. in 
Theureau et al. (1997a).}
\begin{center}
\begin{tabular}{ccccccc}
$T$\tablenotemark{a} & $(M/L_B)_{\rm disk}$\tablenotemark{b} 
& $y_{\rm HI}$\tablenotemark{c} & $y_{\rm tot}$\tablenotemark{d} 
& $M/L(g r_0)$\tablenotemark{e} \\ \tableline
1 & 1.44 & 0.03 & 0.05 & 15.8 \\
2 & 1.37 & 0.04 & 0.06 & 15.2 \\
3 & 1.08 & 0.06 & 0.08 & 14.3 \\
4 & 0.96 & 0.07 & 0.09 & 12.7 \\
5 & 0.87 & 0.09 & 0.11 & 12.6 \\
6 & 0.64 & 0.10 & 0.12 & 9.9 \\
7 & 0.64 & 0.12 & 0.12 & 8.8 \\
8 & 0.76 & 0.14 & 0.14 & 13.2 \\
\end{tabular}
\end{center}
\tablenotetext{a}{ morphological type}
\tablenotetext{b}{disc mass-to-light ratio}
\tablenotetext{c,d}{HI and total gas masses}
\tablenotetext{e}{mass-to-luminosity ratio inside TF measuring radius}
\end{table}

Such values are somewhat larger than what is obtained from the double
galaxy sample by Karachentsev, and a work is in progress to identify
the reason for the inconsistency (Teerikorpi, in prep.). Preliminary
results, taking into account the incompleteness in large separation
distances, suggest that Karachentsev's sample gives evidence for $M/L \approx$
30 for spiral-spiral pairs.

\section{Masses of the Virgo and Perseus-Pisces superclusters and
the Tolman-Bondi model}

We have studied the Perseus-Pisces supercluster (Hanski et al., 1999)
using Malmquist-bias corrected TF distances and the Tolman-Bondi
solutions. Virial masses indicate $M/L =$ 200 -- 600 for the largest PP
clusters (Fig. 1). 
Using matter distribution with an excess density $\propto r^{-2}$,
where $r$ is distance from PP, and a void between PP and the Local Group, 
we get a good fit between Tolman-Bondi model and KLUN data points. If the
$M/L$ values are valid elsewhere in the Universe, we obtain $\Omega_0 =$ 
0.1 -- 0.3. The difference between zero and non-zero cosmological constant 
was found negligible in the Tolman-Bondi calculations. 
\begin{figure}
\plotfiddle{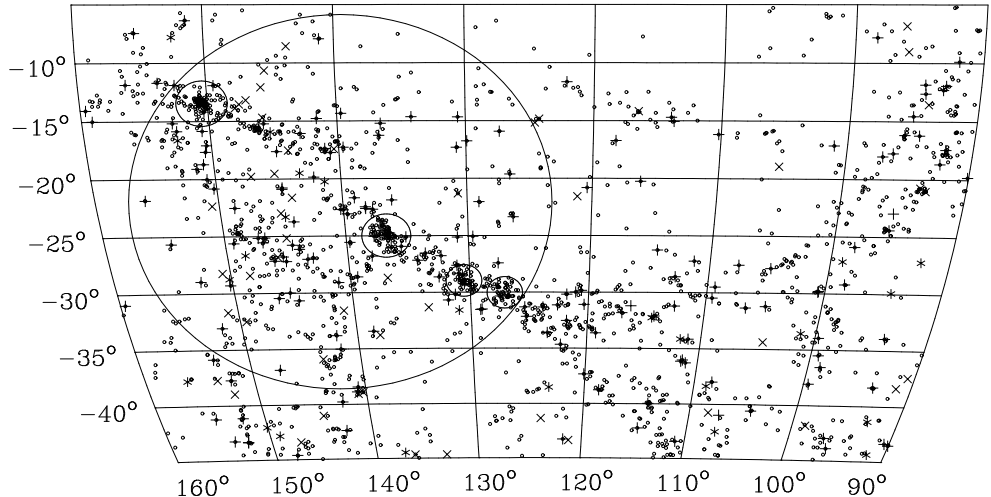}{3in}{270}{80}{80}{-330}{430} 
\caption{LEDA (o) and KLUN galaxies ($+$, $*$ and $\times$) in the PP region.
Large circle is the 15~$h^{-1}$~Mpc sphere at 
$(140.2^{o} ,-22.0^{o} )$ 
containing the main concentration. Small circles are the four densest 
clusters, Perseus, A262, 0122+3305 and 
Pisces, from up left to down right. Using virial and other estimators
by Heisler et al. (1985) we calculated $M/L$ ratios of 280--540, 200--390,
250--550, and 260--590 for these four clusters. See Hanski et al. (1999) 
for details.}
\end{figure}

The  applicability of Tolman-Bondi solution even in regions where the luminous
matter distribution is not spherically symmetric has been evidenced in our 
previous and new work on the Virgo supercluster (Teerikorpi et al.\ 1992,
Ekholm et al., 1999). 
This suggests that the dark matter may be more symmetrically distributed.

\section{Linearity of the Hubble law inside fractal galaxy distribution}

We have applied a developed version of the Sandage-Tammann-Hardy test
of the linearity of the Hubble law inside a hierarchical (fractal) 
galaxy distribution using a linear perturbation approximation for
the velocity-distance law (Baryshev et al., 1998).
\begin{figure}
\plotfiddle{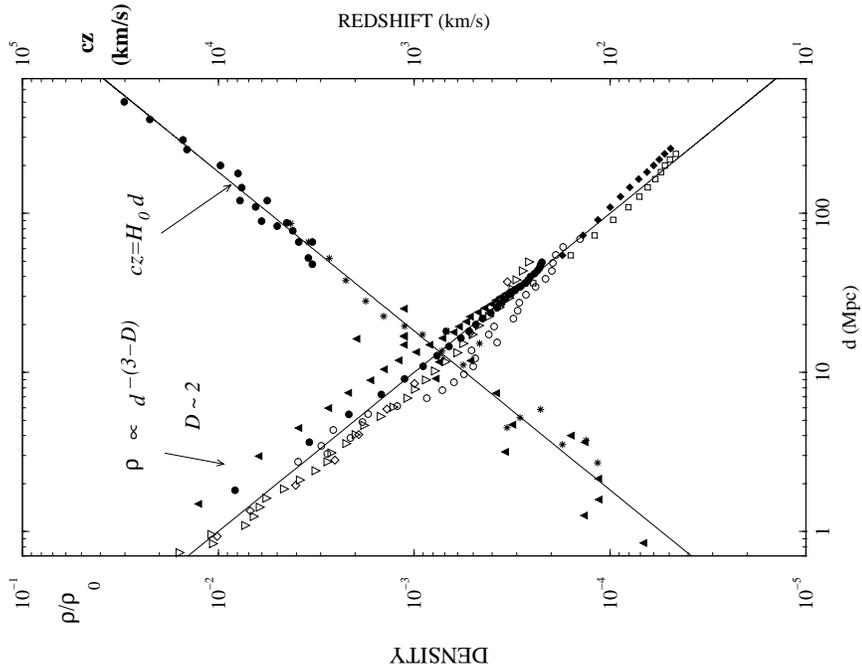}{3.2in}{90}{50}{50}{180}{-18}
\caption{Hubble law vs. galaxy distribution.} 
\end{figure}If fractality with dimension $D\approx 2$ extends up to 200 Mpc,
as suggested by certain studies, including our own using KLUN
(Teerikorpi et al., 1998), a small $\Omega_0$ ($\leq 0.01$) is required 
to produce the good Hubble law. A large $\Omega_0$ is possible if
the dark matter uniformly fills the space. E.g.\ with $\Omega_0=1$,
one derives $\Omega_{\rm dark}=0.99$, and the fractality is restricted
to luminous matter.


\begin{references}
\reference Baryshev, Yu., Sylos Labini, F., Montuori, M., Pietronero, L., \&
Teerikorpi, P. 1998, Fractals, Vol. 6, No. 3, 231
\reference Ekholm, T., Teerikorpi, P., Theureau, G. et al. 
1999, \astap, 347, 99
\reference Ekholm, T., Lanoix, P., Paturel, G., \& Teerikorpi, P., 
1999, \astap, submitted
\reference Hanski, M., Theureau, G., Ekholm, T., \& Teerikorpi, P. 1999, 
\astap, submitted 
\reference Heisler, J., Tremaine, S., \& Bahcall, J. 1985, \apj, 298, 8
\reference Teerikorpi, P., Bottinelli, L., Gouguenheim, L., Paturel, G. 
1992, \astap 260, 17
\reference Teerikorpi, P., Hanski, M., Theureau, G. et al. 
1998, \astap 334, 395
\reference Theureau, G., Hanski, M., Teerikorpi, P. et al. 
1997a, \astap, 319, 435
\reference Theureau, G., Hanski, M., Ekholm, T. et al. 
1997b, \astap, 322, 730
\end{references}
\end{document}